\documentclass[a4paper,12pt]{iopart}

\newcommand{\up}[1][]{_{\uparrow #1}}
\newcommand{\down}[1][]{_{\downarrow #1}}
\newcommand{\LD}{\epsilon}
\newcommand{\SP}{\xi}

\usepackage[latin1]{inputenc}

\usepackage[dvips]{graphicx}
\usepackage{iopams}

\renewcommand{\vec}[1]{\mathbf{#1}}

\begin{document}

\title{FFLO state in 1, 2, and 3 dimensional optical lattices combined with a non-uniform
  background potential}
\author{T.K. Koponen$^{1}$}
\author{T. Paananen$^2$}
\author{J.-P. Martikainen$^{2}$}
\author{M.R. Bakhtiari$^{1}$}
\author{P. T\"{o}rm\"{a}$^{1,3}$}
\address{$^1$ Nanoscience Center, Department of Physics, P.O. Box 35, FI-40014 University of Jyv\"{a}skyl\"{a}, Finland}
\address{$^2$ Department of Physical Sciences, P.O. Box 64, FI-00014 University of Helsinki,  Finland}
\address{$^3$ Laboratory of Physics, Helsinki University of Technology, P.O. Box 1100, 02015 HUT, Finland}
\ead{paivi.torma@hut.fi}

\begin{abstract}
We study the phase diagram of an imbalanced two-component Fermi
gas in optical lattices of 1-3 dimensions, considering the possibilities of the FFLO,
Sarma/breached pair, BCS and normal states as well as phase separation, at
finite and zero temperatures. In particular, phase diagrams with respect
to average chemical potential and the chemical potential difference of the
two components are considered, because this gives the essential information
about the shell structures of phases that will occur in presence of an
additional (harmonic) confinement. These phase diagrams in 1, 2 and 3
dimensions show in a striking way the effect of Van Hove singularities
on the FFLO state. Although we focus on population imbalanced gases,
the results are relevant also for the (effective) mass imbalanced case. We
demonstrate by LDA calculations that various shell structures such as
normal-FFLO-BCS-FFLO-normal, or FFLO-normal, are possible in presence of a
background harmonic trap. The phases are reflected in noise correlations:
especially in 1D the unpaired atoms leave a clear signature of the
FFLO state as a zero-correlation area
(``breach'') within the Fermi sea. This strong signature occurs both
for a 1D lattice as well as for a 1D continuum.
We also discuss the effect of Hartree
energies and the Gorkov correction on the phase diagrams.

\end{abstract}

\submitto{\NJP}
\maketitle


\section{Introduction}
A major experimental breakthrough in the study of ultracold Fermi
gases was the realization of spin-density imbalanced or polarized
Fermi gases \cite{Zwierlein2006a, Partridge2006a, Zwierlein2006c,
  Shin2006a, Partridge2006c, Thomas2006a,Shin2007a}. These experiments are believed to shed
light also on the long standing question of the nature of high-$T_C$
superconductivity \cite{Chen2005a,Giorgini2007a}. In a normal superconductor, the main mechanism for
pairing is the BCS paradigm, however it is not applicable routinely in
the high-$T_C$ systems \cite{Leggett2006a}. Imbalanced Fermi gases
allow the study of states with more exotic pairing, such as the
Fulde-Ferrel-Larkin-Ovchinnikov (FFLO)
\cite{Fulde1964a,Larkin1964a,Larkin1965b} phase and the Sarma or
breached pair (BP) phase \cite{Sarma1963a,Forbes2005a}. These concepts have also been
considered in other fields of physics, such as condensed matter, high
energy and nuclear physics \cite{Casalbuoni2004a,Machida1984a}. It has been shown
experimentally that the imbalanced gas will exhibit 
phase separation by forming a core of BCS superfluid inside a shell of
gas in the normal state in a harmonically trapped system. FFLO-type features have
been predicted to occur in harmonically trapped Fermi gases
\cite{Sheehy2006a, Kinnunen2006a, Machida2006a} as an interface effect \cite{Jensen2007a,Mizushima2007a}. Optical 
lattices are extremely promising in this context, since the lattice enhances the
FFLO-type pairing due to nesting of the Fermi surfaces
\cite{Koponen2007a}. Moreover, optical lattices
allow to manipulate
the effective dimensionality of the system as well as the mobility of
the particles compared to the strength at which they interact
\cite{Bloch2007a}. 
 
Density-density correlations have been used as an indicator for different phases in optical lattices 
for bosonic atoms~\cite{Foelling2005a} and the 
idea is promising also in the fermionic case~\cite{Greiner2005b}. A density-density correlation tells how strongly the atomic densities at different 
positions are correlated.
The reason why density-density correlations can be a useful way to measure different phases in optical 
lattices is that while densities can be very 
similar for different phases, the density-density  correlations can still be markedly different. 
As an example, one can 
mention the Bose-Einstein 
condensate and the Mott insulator phases, which show dramatically different density-density 
correlations~\cite{Altman2004a,Foelling2005a}. In the Mott insulator one can see  
clear correlation peaks in the density-density correlation, but the  density-density correlations vanish (after subtracting the product of average 
densities)
for a Bose-Einstein condensate.
Likewise, it is possible to detect pairing effects in the Fermi gas by measuring density-density correlations
between different  components, as well as antibunching in the density-density correlations in a single component Fermi 
gas~\cite{Rom2006a}.

In this paper, we study phase diagrams for polarized Fermi gases in
optical lattices, taking into account the following states: BCS, BP,
FFLO, normal state, and phase separation into normal and BCS
regions. We account for the effect of the harmonic trap with the use of the local
density approximation and show how shell structures such as
FFLO - normal state will appear in the trap. We also show how the
effective dimensionality of the lattice (3D, 2D or 1D) affects the
phase diagrams and explain the relation to the Van Hove singularities of
the lattices. Furthermore, we study these states 
by measuring the density-density correlation function of the
system. We show that from the structure of the the correlation peaks one can clearly distinguish between 
the spatially modulated FFLO-states, the usual BCS-state, and gain valuable information on the
structure of quasi-particle dispersions. 
Finally we discuss the effects of the Hartree and Gorkov
corrections on the phase diagrams. We focus on the density imbalanced
case, but the results, especially the phase diagrams for fixed
chemical potential difference, are also relevant for the mass
imbalanced case. Different masses could be introduced as different
effective masses originating from different hopping strengths, or by
considering mixtures of fermions of non-equal mass \cite{Wille2007a}.

Note that by 1 and 2 dimensional lattices we mean here: 1D gas in a 1D lattice, and 2D gas in a 2D lattice. 
The first one could be realized by a 3D optical lattice where the confinement is very strong in two directions 
and intermediate in the third one, and the second by strong confinement in one direction and intermediate in two. 
This is different from what is often meant by 1 and 2 dimensional
optical lattices superimposed on a 3D gas, namely that the 1D lattice
forms 2D "pancakes" and 2D lattice forms 1D "tubes". In other words, we study here the actual dimensionality of the lattice, 
not the effective dimensionality of homogeneous space produced by a lattice.

\section{Mean field attractive Hubbard model}
We consider the mean field attractive Hubbard model in the lattice,
\begin{equation}
\label{eq:hamiltonian}
\fl  \widehat{H} = \sum_{\vec{k}} \left(\xi\up[\vec{k}]\hat{c}\up[\vec{k}]^\dagger \hat{c}\up[\vec{k}] + \xi\down[\vec{k}]\hat{c}\down[\vec{k}]^\dagger
    \hat{c}\down[\vec{k}] + \Delta \hat{c}\up[\vec{k}+\vec{q}]^\dagger
  \hat{c}\down[-\vec{k}+\vec{q}]^\dagger
  + \Delta \hat{c}\down[-\vec{k}+\vec{q}]\hat{c}\up[\vec{k}+\vec{q}]\right) - \frac{\Delta^2}{U},
\end{equation}
where we have limited the study to include only the lowest energy
eigenstate of each lattice site, i.e. the lowest band. Here the single particle dispersion is
\begin{equation}
\fl \SP_{\sigma \vec{k}} = 2J_{x}(1 - \cos k_x) + 2J_{y}(1 - \cos k_y) +
2J_{z}(1 - \cos k_z) - \mu_\sigma.
\end{equation}
The interaction term $\Delta
\hat{c}\down[-\vec{k}+\vec{q}]\hat{c}\up[\vec{k}+\vec{q}]$ corresponds
to a plane wave (FFLO) ansatz for the order parameter:
$U\langle\hat{c}\down[\vec{x}]\hat{c}\up[\vec{x}]\rangle = \Delta
e^{2i\vec{q}\cdot\vec{x}}$. The interaction and hopping parameters,
$U$ and $J_i$, are defined as in \cite{Jaksch1998a}. Throughout the
paper, $J$ without an index stands for the largest $J_i$.
Diagonalizing the Hamiltonian with the standard Bogoliubov transformation
yields the quasiparticle energies
\begin{equation}
\fl E_{\pm,\vec{k},\vec{q}} = \frac{\SP\up[\vec{q}+\vec{k}] -
  \SP\down[\vec{q}-\vec{k}]}{2} \pm \sqrt{\left(\frac{\SP\up[\vec{q}+\vec{k}] +
  \SP\down[\vec{q}-\vec{k}]}{2}\right)^2 + \Delta^2},
\end{equation}
and the grand potential of the system is
\begin{equation}
\label{eq:Omega}
\fl \Omega = -\frac{\Delta^2}{U} + \sum_{\vec{k}}\left(\xi\down[-\vec{k}+\vec{q}] +
  E_{-,\vec{k},\vec{q}} - \frac{1}{\beta}\ln \left(\left(1 + e^{-\beta
    E_{+,\vec{k},\vec{q}}}\right)\left(1 + e^{\beta
    E_{-,\vec{k},\vec{q}}} \right) \right)\right),
\end{equation}
where $\beta = 1/k_BT$.

We map out the phase diagrams of the system in section \ref{sec:chemical_diagrams} by minimizing $\Omega$
with respect to $\Delta$ and $\vec{q}$ by keeping the average chemical potential $\mu_{aver}  = (\mu\up +
\mu\down)/2$ and the difference between the chemical potentials,
$\delta\mu = \mu\up - \mu\down$, fixed. We have also studied phase diagrams
for fixed filling factors $n\up$ and $n\down$, in section \ref{sec:density_diagrams}. In this case the
relevant thermodynamical quantity to minimize is the Helmholtz free
energy, which is related to the grand potential by $F = \Omega +
\mu\up n\up + \mu\down n\down$. The major difference between these two
schemes is that with fixed chemical potentials the BCS state cannot
support a finite polarization at zero temperature, and BP is not
stable, but with fixed densities BCS is essentially a special case of
the BP state and can have different numbers of the different spin
components. See e.g. \cite{Koponen2006b} for a more detailed discussion on
the stability of different phases in these two situations.

\section{Phase diagrams}
\label{sec:chemical_diagrams}
Changing the difference of the chemical potentials, $\delta\mu =
\mu\up - \mu\down$, changes the stable phase. At $T=0$, when $\delta\mu =
0$, the result is the standard BCS-state, with $\Delta > 0$ and
$\vec{q} = 0$ and with equal densities in the (pseudo) spin components. As
$\delta\mu$ is increased beyond the Clogston limit \cite{Clogston1962a},
$\sqrt{2}\Delta_0$, where $\Delta_0$ is the energy gap at $\delta\mu = 0$,
the gas switches to a polarized FFLO state, with some finite
$\vec{q}$, which is found by minimizing (\ref{eq:Omega}). This
transition is of the first order. As $\delta\mu$ is increased further,
$\Delta$ approaches zero and $|\vec{q}|$ grows, until the gas undergoes a
second order phase transition to a polarized normal gas. 

\begin{figure}
\center
\includegraphics[width=0.6\textwidth]{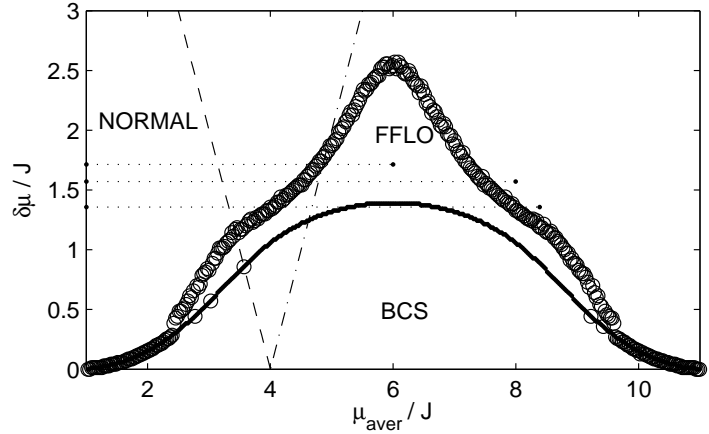}
\caption{The phase diagram of a two-component Fermi gas in a 3D lattice, at zero
  temperature, as function of the average chemical potential
  $\mu_{aver}$, and the difference $\delta\mu$. The dashed line
  (\dashed) shows where $\mu\up = 4J$, that is, the place of the Van
  Hove singularity for the majority component.
The dash-dotted line (\chain) shows where $\mu\down = 4J$, which is
the minority component singularity. Here $J$ is the hopping amplitude. The solid
line shows the calculated data points for the BCS-FFLO (in some points
BCS-normal gas) phase boundary and the circles show the FFLO-normal
gas phase boundary. The dotted (\dotted) horizontal lines correspond
to the shell structures in figure \ref{fig:gap_profiles}. The phase diagram was obtained by minimizing the
grand potential (\ref{eq:Omega}) in each point of the diagram.}
\label{fig:3D_diagram_T0}
\end{figure}
\begin{figure}
\center
\includegraphics[width=0.6\textwidth]{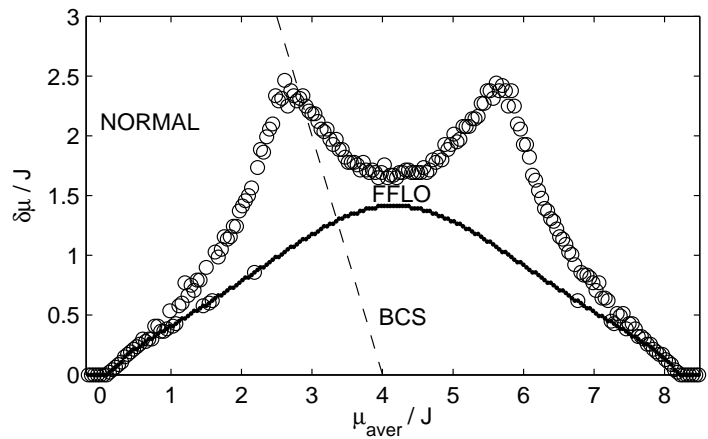}
\caption{The phase diagram of a two-component Fermi gas in an effectively 2D lattice, at zero
  temperature. The dashed line (\dashed) shows where $\mu\up =
  4J$, the Van Hove singularity for the majority component. The calculated data points at the phase boundaries are marked
  so that solid line is BCS-FFLO and circles represent FFLO-normal gas.}
\label{fig:2D_diagram_T0}
\end{figure}
\begin{figure}
\center
\includegraphics[width=0.6\textwidth]{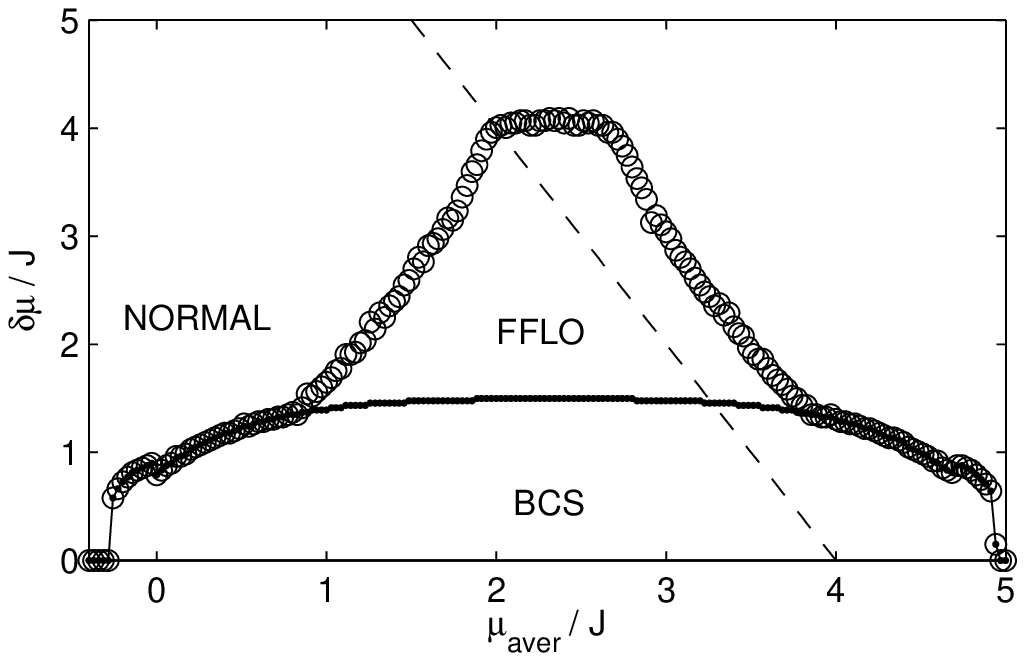}
\caption{The phase diagram of a Fermi gas in an effectively 1D lattice, in zero
  temperature. The dashed line (\dashed) shows where $\mu\up =
  4J$, the Van Hove singularity for the majority component. The calculated phase boundaries are shown 
  so that solid line is BCS-FFLO and circles represent FFLO-normal gas.}
\label{fig:1D_diagram_T0}
\end{figure}

Some typical phase diagrams as functions of $\mu_{aver}  = (\mu\up + \mu\down)/2$
and $\delta\mu$ are given in figures
\ref{fig:3D_diagram_T0}-\ref{fig:more_finite_T_diagrams}. In some of
the diagrams, some of the circles representing the FFLO - normal state
phase boundary have fallen onto the boundary between FFLO and
BCS. This is an artefact of the numerical method employed in the calculations.

\subsection{Van Hove singularity and dimensionality}
The phase diagrams \ref{fig:3D_diagram_T0}-\ref{fig:1D_diagram_T0}
show interesting behaviour regarding the shape of the phase boundary
between the FFLO
and normal states. The reason why FFLO can sustain a finite polarization,
i.e. Fermi surfaces of unequal size, and still be favorable compared
to the normal state, is that it allows the Fermi surfaces to be
partially matched. In FFLO, Cooper pairs have a finite momentum
$2\vec{q}$, which effectively means a relative displacement of the
Fermi surfaces by an amount $\vec{q}$ in order to match them. In 3D and 2D this nesting of the Fermi surfaces is optimal
around the Van Hove singularity, i.e. when the Fermi surfaces touch
the edge of the first Brillouin zone, because the shape of the surface
at such densities is octahedral in 3D and square in 2D. In the
non-interacting case this happens for component $\sigma$ when
$\mu_\sigma = 4J$. Displacing two
octahedra or squares so that their corners connect actually allows to
connect an area, optimally half of the minority Fermi surface, instead of just one point, as in the case of spherical
Fermi surfaces. For this reason the area occupied by FFLO in the phase
diagram is more dominant than observed for spherically symmetric
systems. Moreover, as our results show, the Van Hove singularities
lead to striking features in the FFLO-normal state phase boundary.

The phase boundary between FFLO and normal state shows special
features at the points where the chemical potential of the majority
component, $\mu\up = \mu_{aver}  + \delta\mu/2$ has the value $4J$. In
the 3D diagram, figure \ref{fig:3D_diagram_T0}, this produces a kink
at $\mu_{aver} \approx 3.4J$. The phase boundary continues to grow,
showing another slight change of shape around the minority component
Van Hove singularity. In 2D (figure
\ref{fig:2D_diagram_T0}), the critical value of $\delta\mu$ has a
pronounced maximum at the singularity at $\mu_{aver} \approx 2.9J$,
after which it decreases. The singularity in 2D is stronger than in 3D
and the minority component does not reach its Van Hove singularity
before half filling.
In 1D this also produces a clear feature at $\mu_{aver}  \approx 2J$, where
the phase boundary essentially becomes horizontal, see figure
\ref{fig:1D_diagram_T0}. 

In our calculations we have taken the lattice height to be $2.5 E_R$
and in the 2D and 1D calculations we have taken the lattice height in
the orthogonal directions to be $10 E_R$. We have used $6$ for the
mass number of the atoms, corresponding to Lithium. Our calculations
are in the intermediate coupling regime, with $U/J < 6$. We have also
checked our results in the weak coupling BCS limit and found no qualitative
difference to the results presented here. The explicit coupling
strengths ($U/J$) used are $-3.7$ in 3D, $-3.3$ in 2D and $-3.2$ in
1D. Note that all calculations are performed for the full 3D system,
just with the higher lattice heights in the orthogonal
directions. Therefore the calculations take into account the small but
finite tunneling between the one or two dimensional systems. 

\subsection{Finite temperature}
\begin{figure}
\begin{minipage}{0.495\textwidth}
\centering
a
\includegraphics[width=\textwidth]{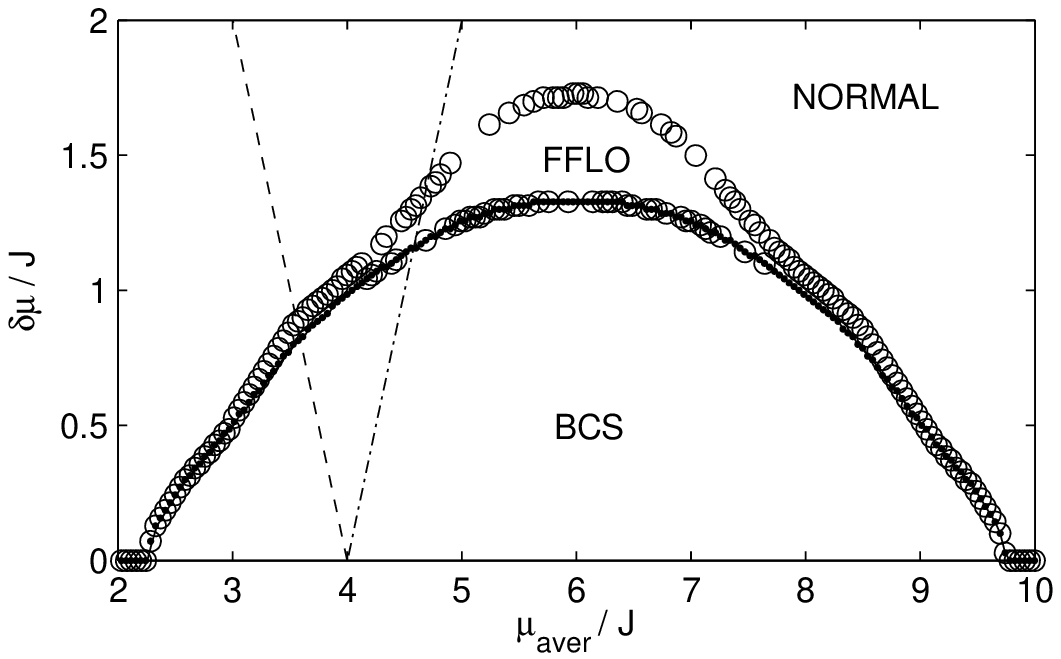}
\end{minipage}
\begin{minipage}{0.495\textwidth}
\centering
b
\includegraphics[width=\textwidth]{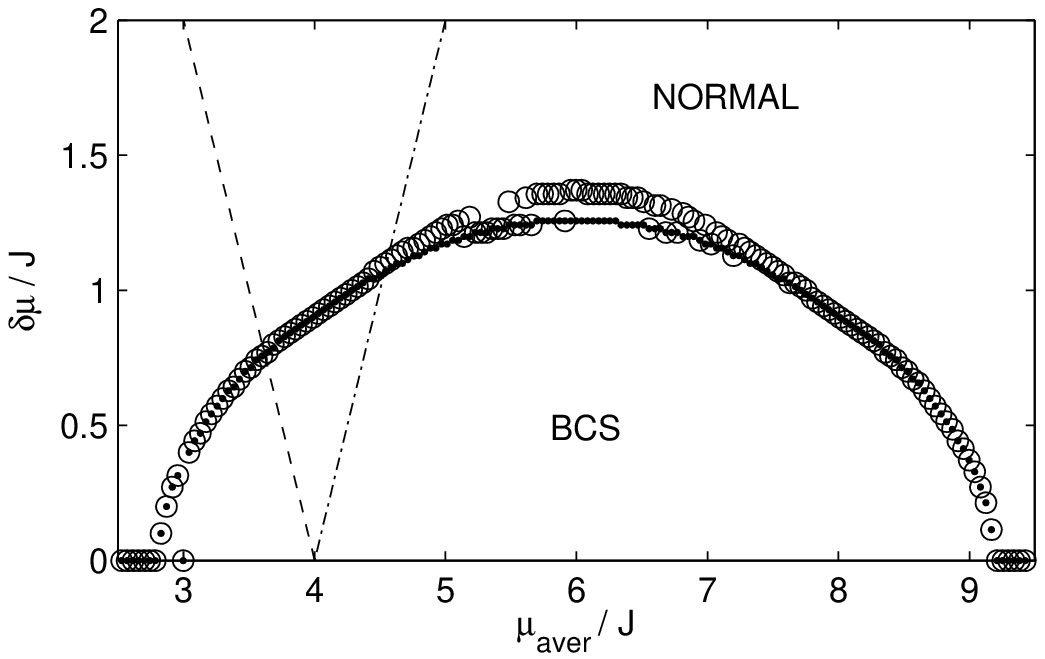}
\end{minipage}
\begin{minipage}{0.495\textwidth}
\centering
c
\includegraphics[width=\textwidth]{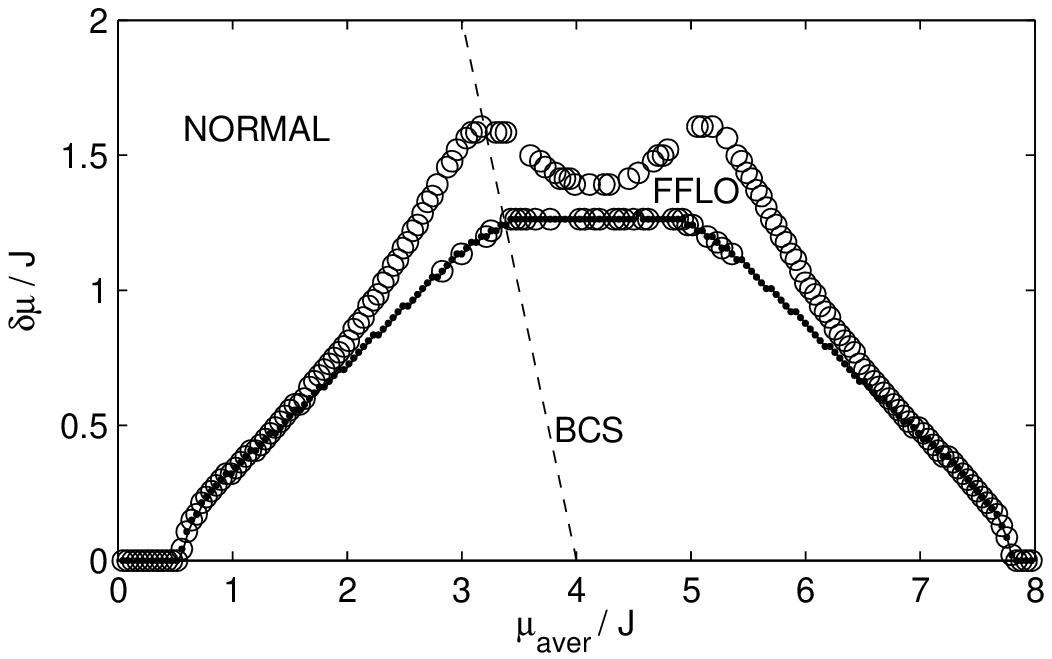}
\end{minipage}
\begin{minipage}{0.495\textwidth}
\centering
d
\includegraphics[width=\textwidth]{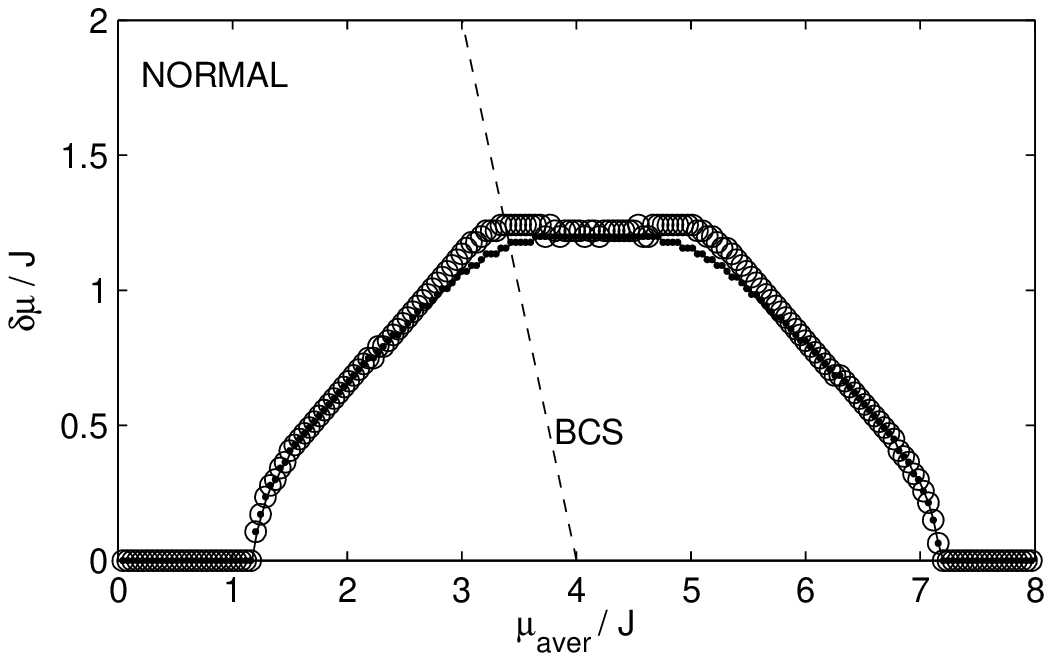}
\end{minipage}
\caption{Finite temperature phase diagrams in 3D and 2D lattices: (a)
  3D at $k_B T / J \approx 0.1$ (or $10$ nK), (b) 3D at $k_B T / J
  \approx 0.2$ (or $20$ nK),
  (c) 2D at $k_B T / J \approx 0.1$ (or $10$ nK), and (d) 3D at $k_B T / J \approx
  0.2$ (or $20 nK$). The dashed line (\dashed) always shows the majority Van Hove
  singularity, where $\mu\up = 4J$, and the dash-dotted line (\chain)
  shows the minority singularity, with $\mu\down = 4J$ (only reached
  in 3D). The calculated data points at the phase boundaries are
  solid: BCS-FFLO and circles: FFLO - normal state. The temperatures in nK
  correspond to $J_x = 0.07$ $E_R$.}
\label{fig:3D_2D_diagram_finite_temperature}
\end{figure}

\begin{figure}
\begin{minipage}{0.495\textwidth}
\centering
a
\includegraphics[width=\textwidth]{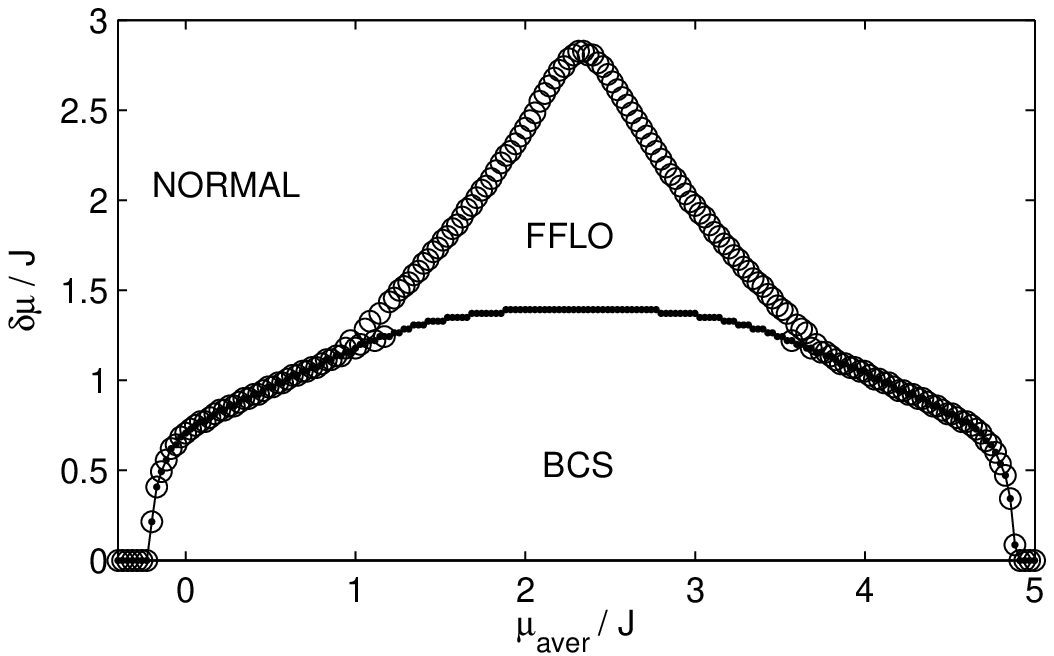}
\end{minipage}
\begin{minipage}{0.495\textwidth}
\centering
b
\includegraphics[width=\textwidth]{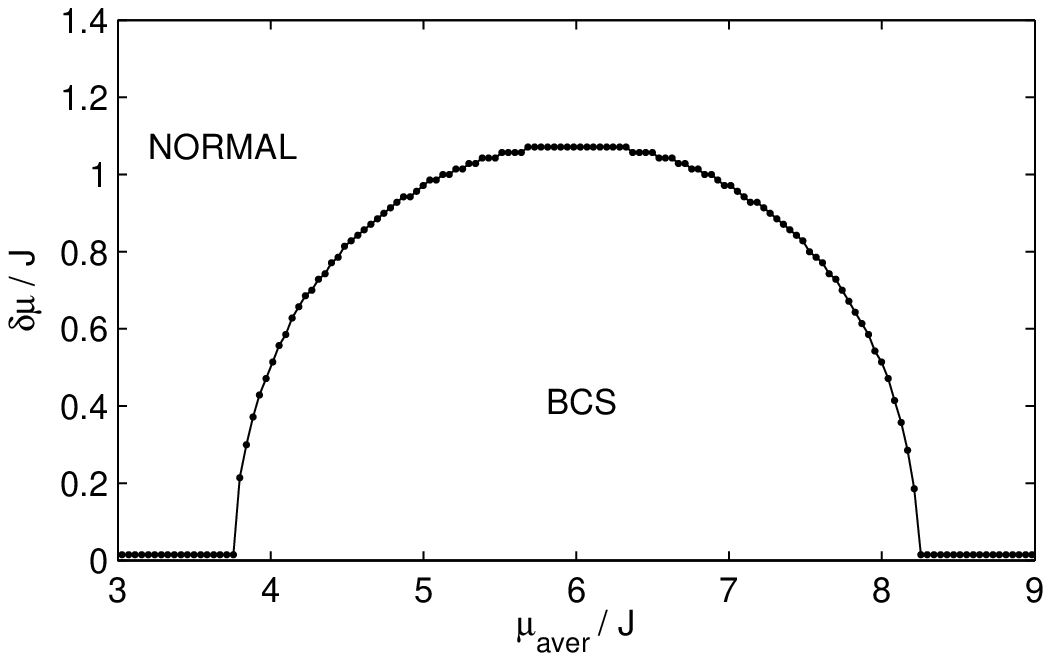}
\end{minipage}
\begin{minipage}{0.495\textwidth}
\centering
c
\includegraphics[width=\textwidth]{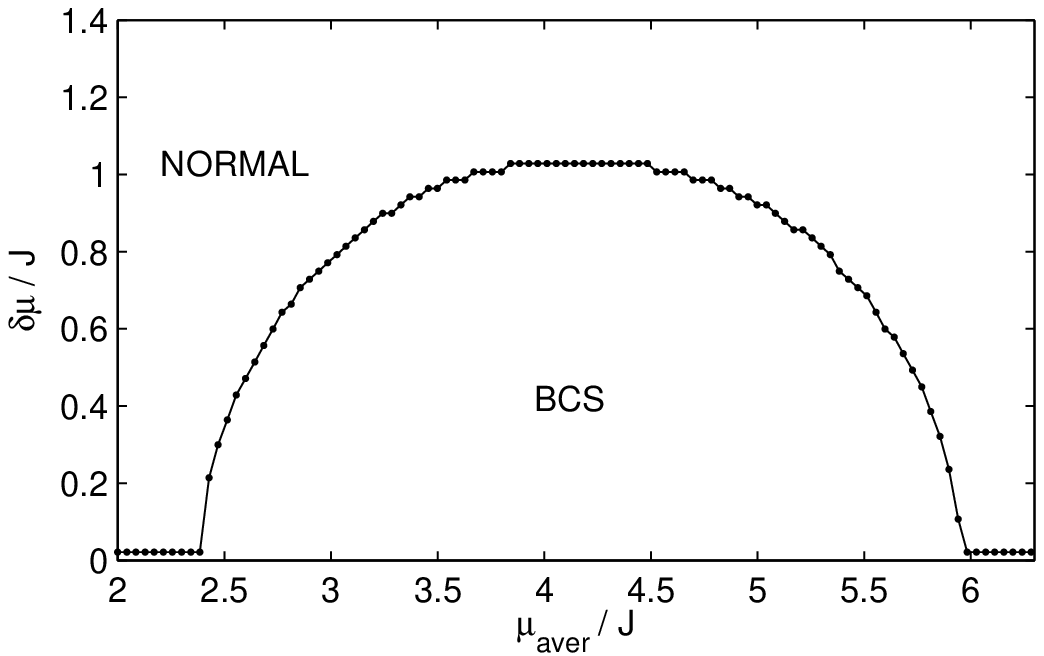}
\end{minipage}
\begin{minipage}{0.495\textwidth}
\centering
d
\includegraphics[width=\textwidth]{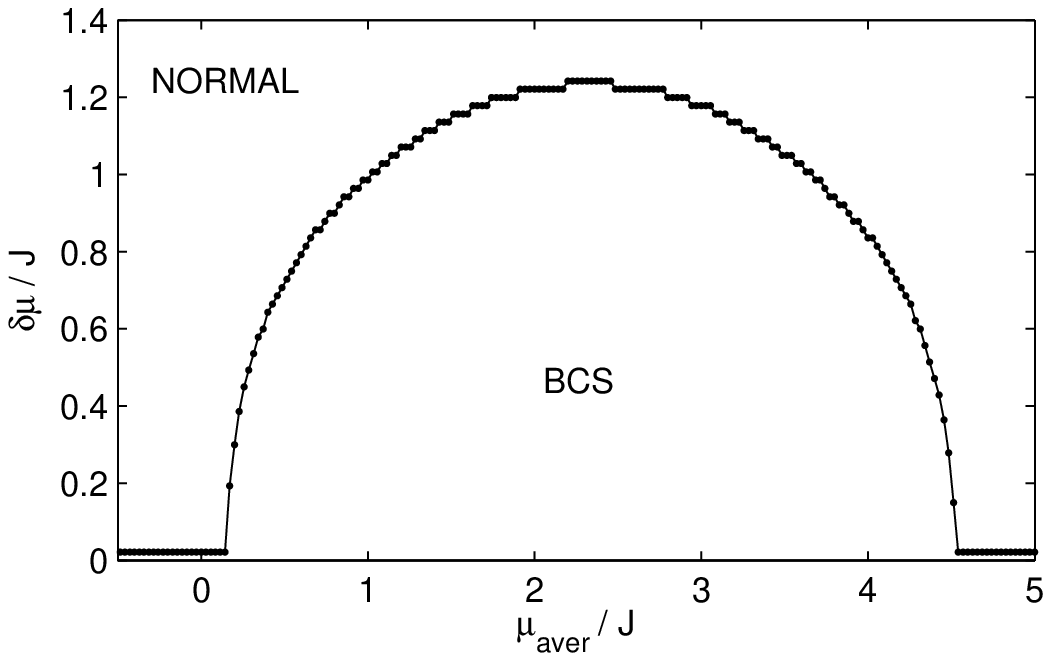}
\end{minipage}
\caption{Finite temperature phase diagrams: (a) 1D at $k_B T / J
  \approx 0.2$ (or $20$ nK), (b) 3D at $k_B T / J \approx 0.4$ (or $40$ nK), (c) 2D at $k_B T /
  J \approx 0.4$ (or $40$ nK), and (d) 1D at $k_B T / J \approx 0.4$ (or
  $40$ nK). In (a), the flat top (see figure \ref{fig:1D_diagram_T0}) disappears, because
  the maximum $\delta\mu$ is so small that $\mu\up$ does not reach the
  Van Hove singularity. In the high-temperature diagrams, (b)-(d), FFLO and all the features
  related to the dimensionality, have vanished. The temperatures in nK
  correspond to $J_x = 0.07$ $E_R$.}
\label{fig:more_finite_T_diagrams}
\end{figure}

We have studied the phase diagrams also in finite temperature, and our
results indicate that the FFLO area, and features related to it, of
the phase diagram, will gradually disappear with increasing temperature, as shown in figures
\ref{fig:3D_2D_diagram_finite_temperature}-\ref{fig:more_finite_T_diagrams}. In high
temperatures even the effect of dimensionality seems to disappear, and
phase diagrams b-d in \ref{fig:more_finite_T_diagrams} have
no qualitative differences. It should be noted that in lattices the
BCS critical temperature $T_C$ depends on the average filling
factor or average chemical potential, since $T_C$ is directly
proportional to the gap at zero temperature. At half filling $T_C$ is
much higher than at low fillings. At finite temperatures, the BCS state
can also support some polarization, due to thermal fluctuations.

\section{Effect of the harmonic confinement}
\label{sec:trap}
We have used the local density approximation to study effects of a
harmonic trapping potential superimposed on the lattice. Assuming that
the harmonic trap is sufficiently shallow, it is possible to
account for its effects by taking it to be locally constant and letting
the chemical potential vary as a function of position, the standard
local density approximation (LDA). This means that
we take the system to consist of cubical 
regions of $N_{grid}^3$ lattice sites, where we approximate the local
chemical potential for each component to be $\mu_\sigma'(r) =
\mu_\sigma - V(r)$, where $r$ is the distance from the center of the
trap to the center of the cube, $V(r)$ is the trapping potential, and
$\mu_\sigma$ is the global chemical potential for component $\sigma$,
which is constant throughout the system. Because the system is
approximated with an infinite lattice inside each region, the value of
$N_{grid}$ can be chosen freely and does not have any physical implications.

\begin{figure}
\centering
\begin{minipage}{0.495\textwidth}
\centering a
\includegraphics[width=\textwidth]{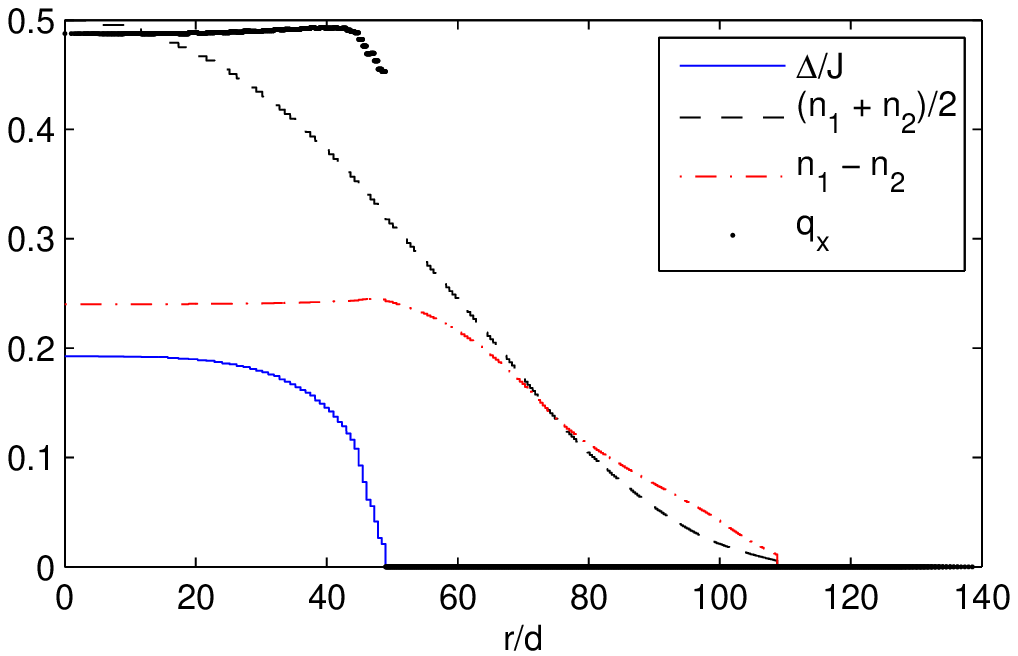}
\end{minipage}
\begin{minipage}{0.495\textwidth}
\centering b
\includegraphics[width=\textwidth]{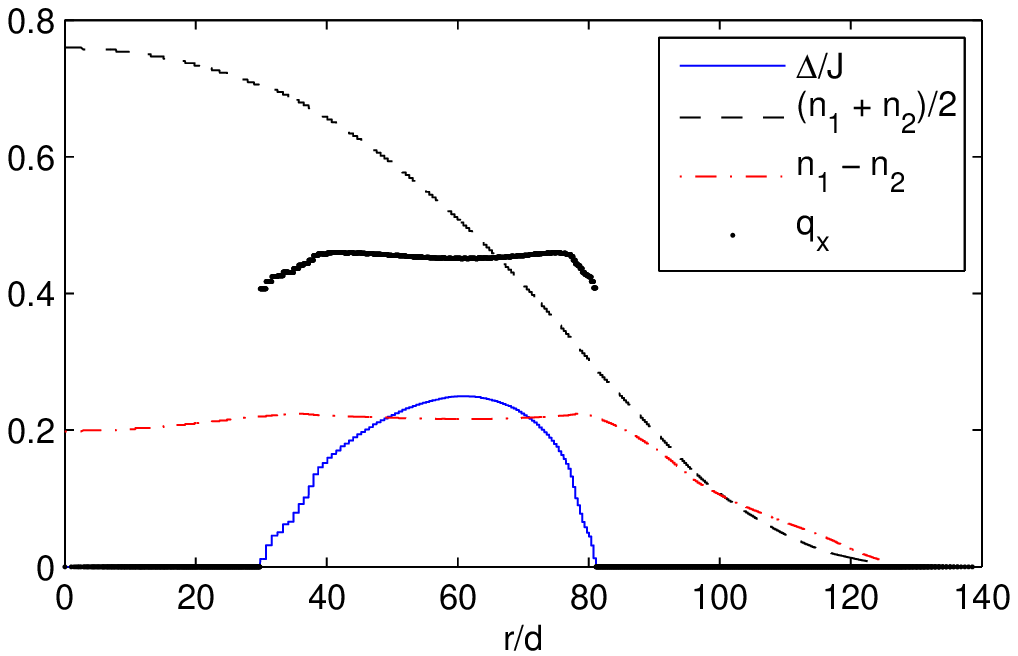}
\end{minipage}
\begin{minipage}{0.495\textwidth}
\centering c
\includegraphics[width=\textwidth]{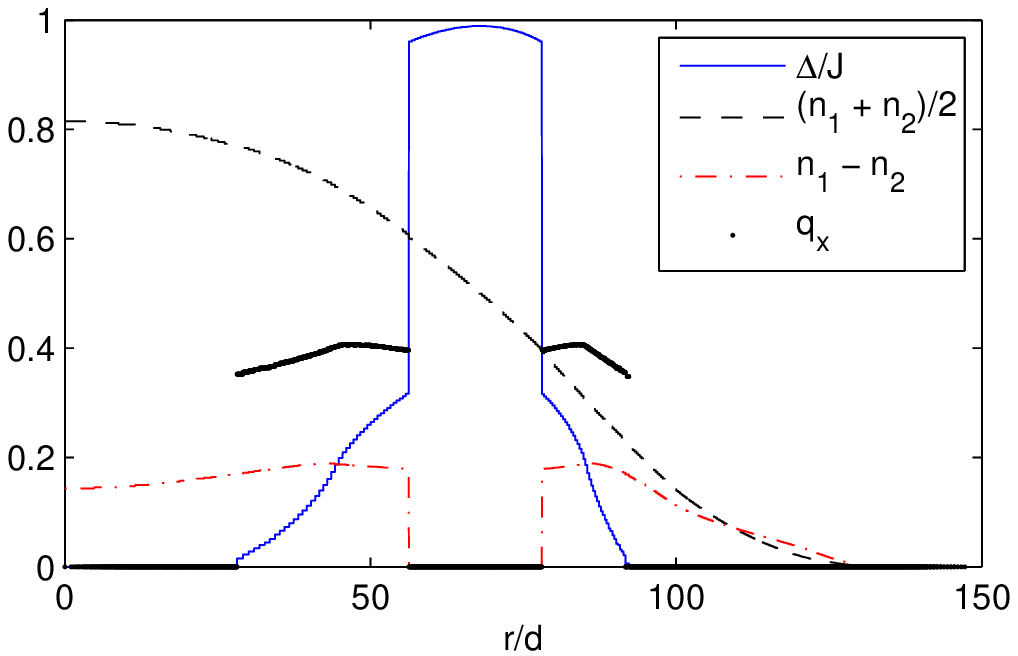}
\end{minipage}
\begin{minipage}{0.495\textwidth}
\centering d
\includegraphics[width=\textwidth]{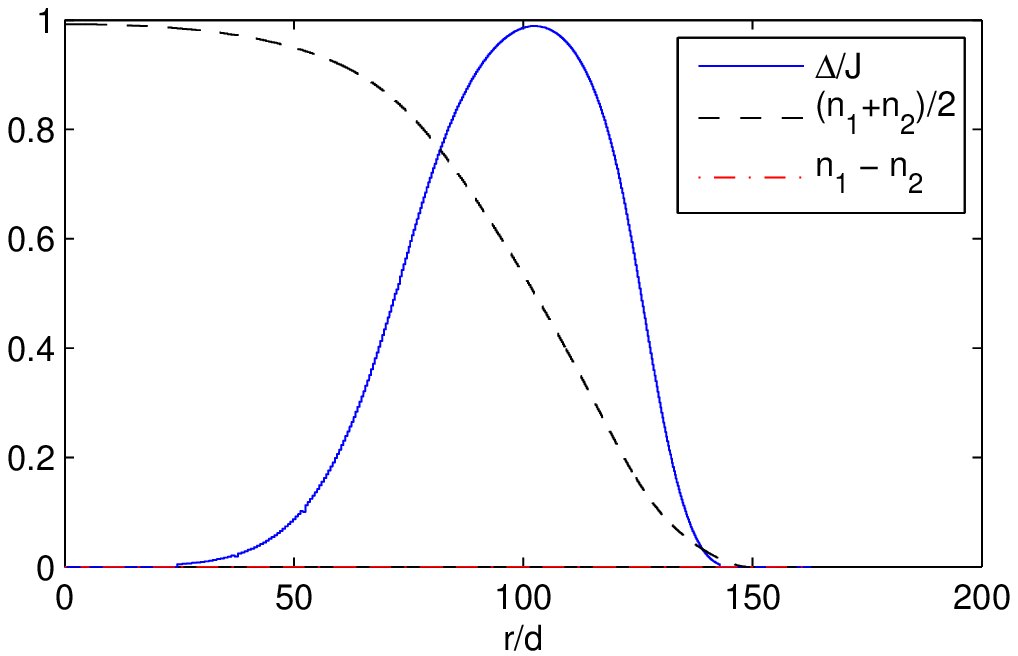}
\end{minipage}
\caption{Gap ($\Delta/J$), average density ($(n_1 + n_2)/2$), and density
  difference ($n_1 - n_2$) profiles for a 3D lattice in a harmonic trap with
  LDA, together with the FFLO wave vector
  $\vec{q}$. The frequency of the harmonic trap is chosen to be $120$ Hz and
  the wavelength of the lattice $d$, $515$ nm. Shown are four
  different shell structures: (a) with $\delta\mu/J \approx 1.7$ and
  the density in the center of the trap at half filling, the system
  forms a polarized core in the FFLO state, surrounded by a shell of
  polarized normal gas. In (b), $\delta\mu/J \approx 1.6$, but there
  are more atoms overall in the system, giving a higher density in the
  center of the trap, which leads to a core of normal gas to form inside a
  shell of FFLO. Figure (c) shows an even more intriguing shell
  structure, corresponding to $\delta\mu/J \approx 1.4$: normal gas -
  FFLO - BCS - FFLO - normal gas. Note that this structure is
  dramatically reflected in the density difference. In (d), the system is unpolarized,
  but the density in the center of the trap is so high that a core of
  normal gas is formed inside a shell of BCS superfluid. All the plots
correspond to horizontal lines in figure \ref{fig:3D_diagram_T0}. The
overall polarizations $P = (N\up - N\down)/(N\up + N\down)$ are: (a)
0.44, (b) 0.27, (c) 0.13, and (d) 0.}
\label{fig:gap_profiles}
\end{figure}

In this scheme, reading the effect of the trapping potential from the
phase diagrams computed with respect $\mu_{aver} $ and $\delta\mu$ is
exceedingly simple: because both of the components see the same trapping
potential, $\delta\mu$ does not depend on $r$, and any situation with given numbers of atoms in each component
corresponds to a horizontal line in the appropriate phase
diagram. Since the maximum number of identical fermions in the ground state of
each lattice site is 1 for each component in the gas, the order
parameter $\Delta$ goes to zero as the filling factor of either of the
components approaches zero or one. This implies that with sufficiently high
total particle numbers the region in the center of the trap is not
superfluid, but a normal gas, surrounded by a superfluid shell, which
is again surrounded by a shell of normal gas \cite{Helmes2007a,Chen2007a}. Because at zero
temperature, with any filling besides close to an empty or a full
lattice, FFLO is always between the BCS and the normal states in the
phase diagram, there is a shell of FFLO between the BCS and normal
regions in the trap. This is demonstrated in the gap profiles in
figure \ref{fig:gap_profiles}.

\section{Density-density correlations}
Here we compute the density-density correlation functions after free expansion. Correlation functions are defined by~\cite{Altman2004a}
\begin{eqnarray}
\label{eq:kormaar}
 G_{\uparrow  \downarrow }(\vec{r},\vec{r'},t)&=\langle \hat n_{\uparrow }(\vec{r},t)\hat n_{\downarrow }(\vec{r'},t)\rangle-
\langle \hat n_{\uparrow }(\vec{r},t)\rangle\langle \hat n_{\downarrow }(\vec{r'},t)\rangle\\
&=\langle \hat \Psi_{\uparrow }^{\dagger}(\vec{r},t)\hat \Psi_{\uparrow }(\vec{r},t)\hat \Psi_{\downarrow }^{\dagger}(\vec{r'},t)\hat \Psi_{\downarrow }(\vec{r'},t)\rangle-
\langle \hat n_{\uparrow }(\vec{r},t)\rangle\langle \hat n_{\downarrow }(\vec{r'},t)\rangle, \nonumber 
\end{eqnarray}
where $\uparrow $ and $\downarrow $ are the component indices and $t$ is time.
In the above formula we have subtracted the term with mean atomic densities, since often it is easier to focus on fluctuations.
Because the density-density correlations after 
the free expansion reflect  correlations in momentum space at $t=0$, correlations can be computed using the 
wave-function in momentum space prior to expansion~\cite{Paananen2008a}. 
Positions in real space after time $t$ are related to $\vec{ k}$-vectors trough $\vec{ r}=\hbar t \vec{k}/m$.
 
In momentum space it turns out that the plane-wave FFLO 
density-density correlations are strongly correlated at points which correspond to momenta $\vec{k}$ and $-\vec{k}+2\vec{q}$, where
$\vec{q}$ is the wavevector associated with the FFLO state. In the BCS
state density-density correlations peak 
similarly, but with 
$\vec{q}=0$. The two-mode FFLO-state i.e. the state where the pairing gap is given by 
$\Delta(\vec{r})=\Delta_0\cos (\vec{q}\cdot \vec{r})$, can also leave 
a clear signature on the density-density correlations.
In the two-mode FFLO-state an $\uparrow$-atom which is in the momentum state 
${\vec{k}}$ is paired with  $\downarrow$-atoms which are in the 
momentum states $-\vec{k}+\vec{q}$ and  $-\vec{k}-\vec{q}$. This gives rise to  
large correlation  peaks when $\vec{k}+\vec{k'}\pm\vec{q}=0$.

\begin{figure}
\begin{tabular}{ll}
\includegraphics{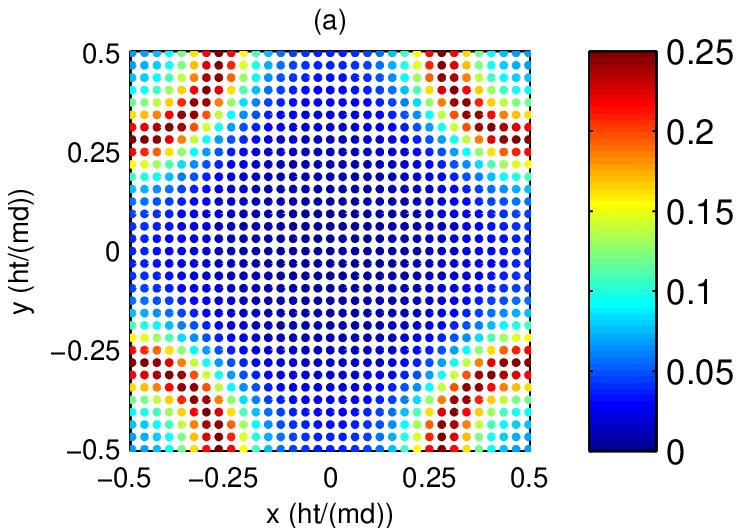} & \includegraphics{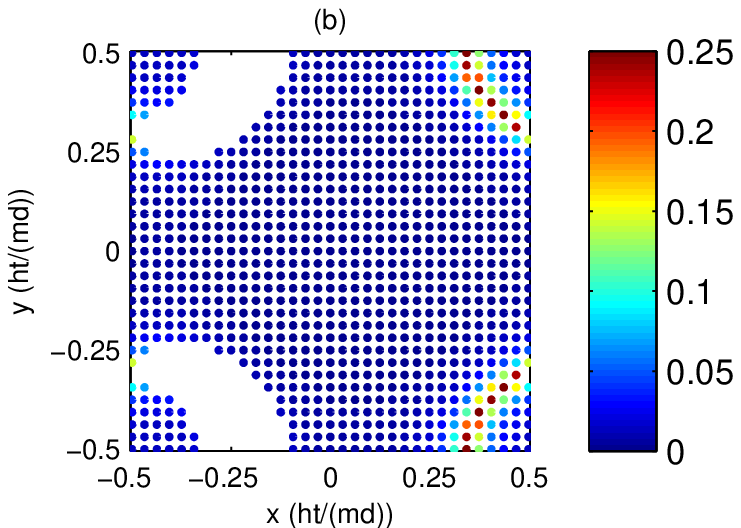} \\
\includegraphics{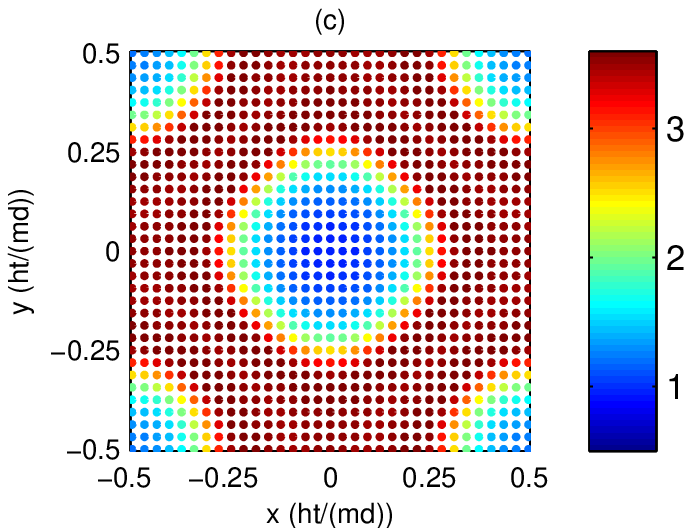} & \includegraphics{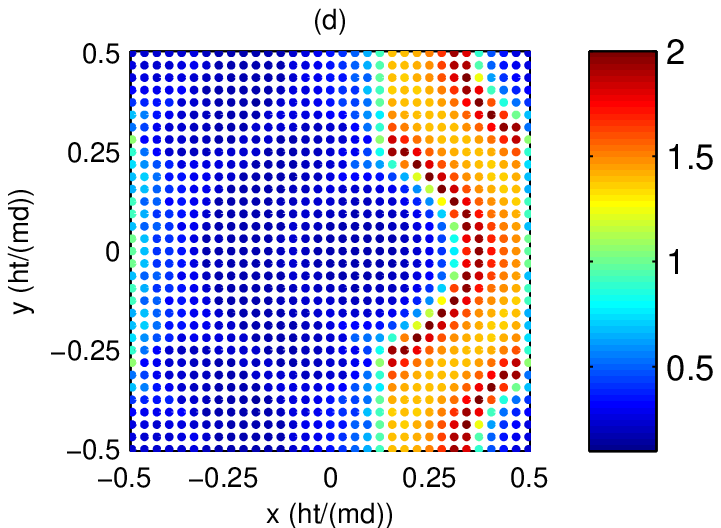}
\end{tabular}
\caption{
Differences between the BCS-state density-density correlations and
the FFLO-state  density-density correlations in a 3D lattice.
In (a) is shown an example of the 
BCS-state density-density correlations in the $z=0$ plane, while in (b) we show an example of 
the FFLO-state density-density correlation in $z=0$ plane.
In (c) and (d) are shown the corresponding signals which have been integrated over $z$.
In (a) and (c) we chose $\vec{r}=-\vec{r'}$ and in 
(b) and (c)  $\vec{r}=-\vec{r'}+\hbar t2\vec{q}/m$. Note that any other
choice would produce zero correlation. All these examples were calculated at zero temperature 
and in (a) and (c)
$P=0.0$, $(n_\uparrow+n_\downarrow )/2=0.55 $, and $\Delta/(2J)=0.49$ and in (b) and (d)
$P=0.168$, $(n_\uparrow+n_\downarrow )/2=0.55 $, $2q_x=0.25(\pi/d)$, $q_y=q_z=0$, and $\Delta/(2J)=0.16$.
Color-coding is such that warm colors imply high peaks and
cold colors low, but in the white areas the correlations vanish identically.
 }
\label{fig:kohinakorrelaatio}
\end{figure}

In figure~\ref{fig:kohinakorrelaatio} we demonstrate the difference between the BCS density-density 
correlations and the FFLO density-density correlations at zero temperature.
In figure (a) we have plotted an example of the BCS-state
density-density correlation between the components at the $z=0$ plane 
while in figure (b)  
we show on example of the FFLO density-density correlation between
components, again at the same plane. 
As one can clearly see, the correlations are very different and the difference arises from
two reasons. First, in the FFLO-state the density-density correlation peaks have been effectively shifted by ${2\vec{ q}}$. 
Second, in the FFLO-state density-density correlation there are areas in the momentum space
where correlation peaks vanish. The reason for this ``breach'' is that one quasiparticle dispersion has changed sign
in the peakless region. Physically 
this means that these areas are populated only by normal atoms, and there are no pairs to give rise
to correlation peaks. The height of the correlation peaks contains information on the underlying pair wave-function
$\sim u_{\vec{k}}v_{\vec{k}}$~\cite{Paananen2008a}. In the weakly interacting BCS limit this function is strongly peaked
at the Fermi surface, but it becomes broad and featureless in the BEC
limit. In (c) and (d), we show that the difference between the BCS- and FFLO-state 
density-density correlations can persist even for integrated correlation signals
\begin{equation}
\fl C_{\uparrow,\downarrow}(x,y)=\int\, dz\,
G_{\uparrow,\downarrow}\left(x,y,z,-x+\hbar t2q_x/m,-y+\hbar
  t2q_y/m,-z+\hbar t2q_z/m\right),
\end{equation}
although integration obviously smooths out some features, especially
the complete ``breach'' is not visible. At non-zero temperatures, sharp areas without correlation peaks disappear with increasing temperatures. 
However, the shift in the positions of the peaks persists even at non-zero temperatures~\cite{Paananen2008a}.
 
\begin{figure}
\begin{tabular}{ll}
\includegraphics{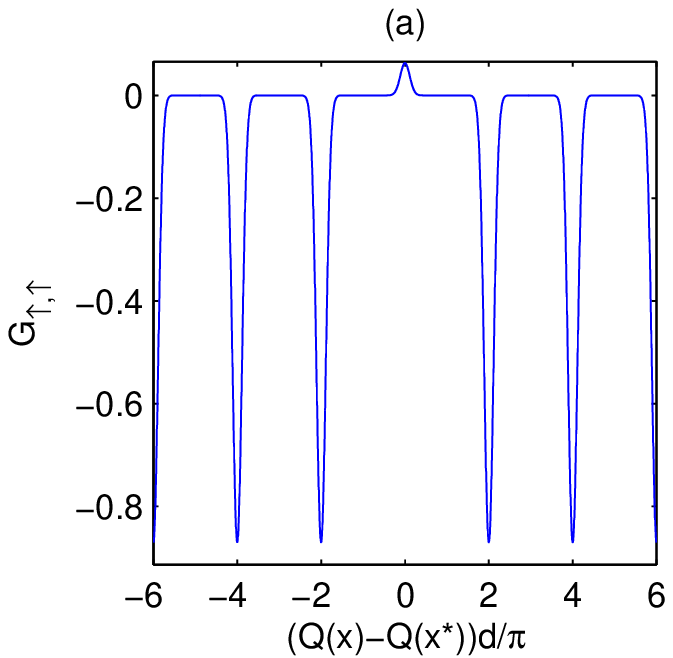} & \includegraphics{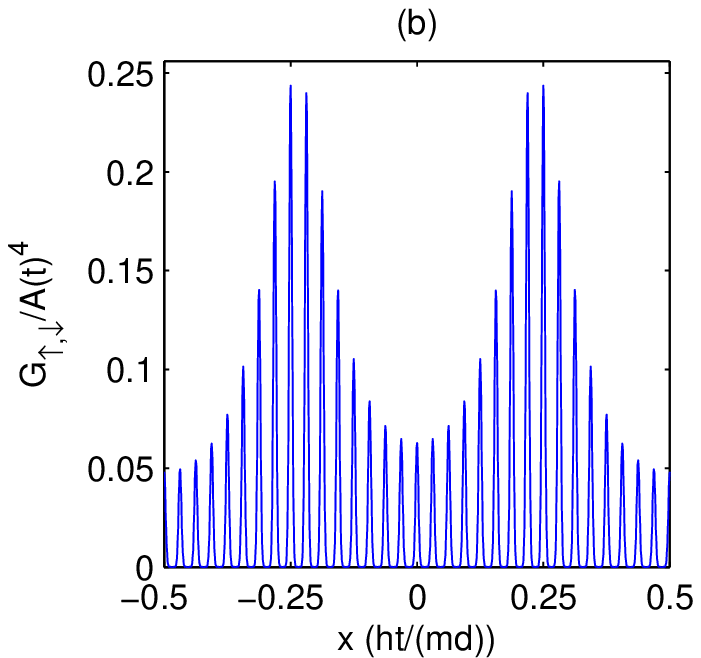} \\
\includegraphics{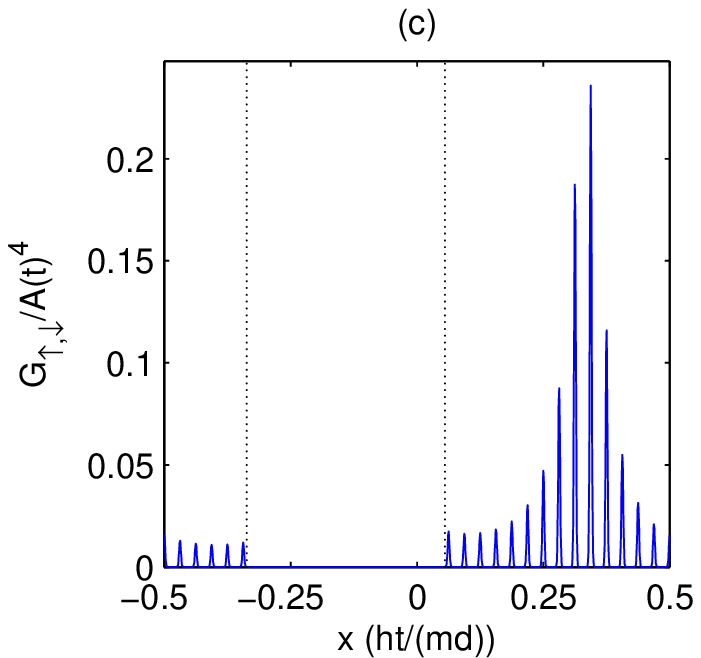} & \includegraphics{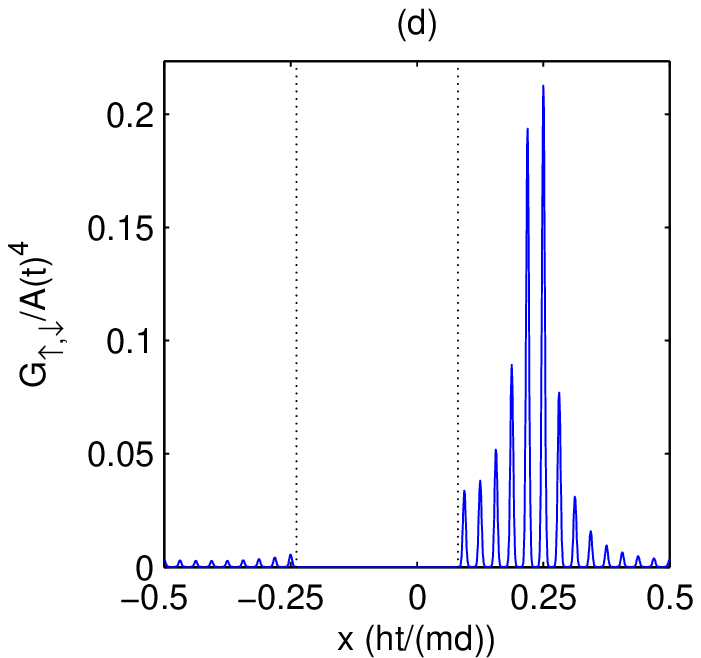}
\end{tabular}
\caption{
1D lattice density-density correlations. In (a) we demonstrate the antibunching effects of the BCS density-density correlation of
a single component  in a 1D lattice.
In (b) is shown the BCS-state density-density correlation between components and 
in (c), we show the FFLO-state density-density correlation between
components and (d) the FFLO-state density-density correlation between
the components with low average filling fraction. In (c) and (d) dotted lines indicate the gapless regions.
In (a) and (b) the polarization $P=0$, $(n_\uparrow+n_\downarrow )/2=0.41$ and $\Delta/(2J)=0.53$.
In (c) the polarization $P=0.48$, $(n_\uparrow+n_\downarrow )/2=0.40$, $2q_x=0.40 (\pi/d)$ and $\Delta/(2J)=0.19$.
In (d) the polarization $P=0.91$, $(n_\uparrow+n_\downarrow )/2=0.20$, $2q_x=0.37 (\pi/d)$ and $\Delta/(2J)=0.041$.
In (a), $Q(x)=m x/(\hbar t)$ and we have chosen $x=0$.
In (b), we have chosen $x+x'=0$ and in (c) and (d) we have chosen
$x+x'-2\hbar t q_x/m=0$. For other choices, correlations vanish. The distance
between the peaks in figures 8b-d, as well as in figure 7, reflect the
discreteness of the finite size lattice.}
\label{fig:kohinakorrelaatio2}
\end{figure}

The difference between density-density correlations of the FFLO-state
and the BCS-state can be seen even more clearly when the lattice is one dimensional (1D) 
i.e. when tunneling strengths in y- and z-directions vanish. In
figure~\ref{fig:kohinakorrelaatio2} we show an example of density-density correlations in a 1D lattice.
There (a) demonstrates the antibunching in the BCS-state
density-density correlation of a single component.
The result is similar to the one in the ideal Fermi gas \cite{Rom2006a}. 
However, in the superfluid BCS state a "bunching" peak appears at
$Q(x)-Q(x')=0$, where $Q(x)=m x/(\hbar t)=\vec{k}$, where $\vec{k} $ is a lattice momentum.
This peak is absent for the ideal Fermi gas.
In (b), we demonstrate the BCS-state density-density correlation
between the components and 
in (c) we show the FFLO-state density-density correlation between the components.
As one can see from figure (b) the the BCS-state density-density correlation between components is symmetric with respect to $x=0$ and it has 
no regions without correlation peaks.
On the other hand, from figure (c) we can see that the FFLO-state density-density correlation has a region (marked with dashed lines) without correlation peaks and one 
can also see that the FFLO-state density-density correlation is not
symmetric with respect to $x=0$. Note especially that in 1D the region
empty from correlation peaks is not vanished by integration over other
dimensions. In figure (d) we show that this remarkable signature
appears in the low filling case as well, which corresponds to the
continuum 1D system that is of current experimental interest as well.



Interestingly, the phase-separation between the normal gas and a  
paired-state could be visible in the density-density correlations  
between components. This follows from the fact that the  
density-density correlations between components in the normal state  
vanish,
whereas in the paired state the correlations are at their strongest  
around the Fermi momentum. In a lattice superimposed by a trap the  
local density, and therefore the local Fermi momentum, is different in  
different areas of the gas. This may allow to identify spatial phase  
separation and shell structures of normal and paired states from the  
freely expanded cloud where momentum has been mapped into position.

\section{Fixed density phase diagrams}
\label{sec:density_diagrams}
In ultracold gas experiments, the number of particles is often the
fixed quantity, not the chemical potential. Although the trapping
potential makes also fixed chemical potential calculation relevant in
the LDA sense (as discussed in section \ref{sec:trap}), it is of
interest to consider also the fixed particle number case, because in
the case of optical lattices the background potential is a practical
issue which may be eliminated to a certain extent.
We have studied the phase diagrams as a function of polarization
$P = (n\up- n\down)/(n\up + n\down)$ and temperature, i.e. keeping the
filling factors $n\up$ and $n\down$ constant with respect to position.

In this situation, in addition to minimizing the free energy
$F$, it is necessary to solve the chemical potentials from the number
equations:
\begin{eqnarray}
\nonumber N\up &= \sum_{\vec{k}} \left\langle \hat{c}\up[\vec{k}]^\dagger
      \hat{c}\up[\vec{k}]\right\rangle = \sum_{\vec{k}} u_k^2
    f(E_{+,\vec{k},\vec{q}}) + v_k^2f(E_{-,\vec{k},\vec{q}})\\
    N\down &= \sum_{\vec{k}} \left\langle \hat{c}\down[\vec{k}]^\dagger
      \hat{c}\down[\vec{k}]\right\rangle = \sum_{\vec{k}} u_k^2
    f(-E_{-,\vec{k},\vec{q}}) + v_k^2f(-E_{+,\vec{k},\vec{q}}),
\label{eq:number_equations}
\end{eqnarray}
where
\[
\fl u_k^2 = \frac{1}{2}\left(1 + \frac{\xi\up[\vec{k}+\vec{q}] +
    \xi\down[-\vec{k}+\vec{q}]}{2\sqrt{\left(\frac{\xi\up[\vec{k}+\vec{q}] +
      \xi\down[-\vec{k}+\vec{q}]}{2}\right)^2 + \Delta^2}} \right),
\qquad v_k^2 = \frac{1}{2}\left(1 - \frac{\xi\up[\vec{k}+\vec{q}] +
    \xi\down[-\vec{k}+\vec{q}]}{2\sqrt{\left(\frac{\xi\up[\vec{k}+\vec{q}] +
      \xi\down[-\vec{k}+\vec{q}]}{2}\right)^2 + \Delta^2}} \right)
\]
and $f$ is the Fermi function.
This scheme produces the following stable phases:
a polarized superfluid phase, which is the standard BCS phase when
$P=0$ and the BP/Sarma phase when $P>0$, FFLO, and normal state. It is important to note that
when the densities are kept fixed, the BCS state is a special case of
the BP state, with zero polarization. In addition to these phases, we have considered a
phase separated state consisting of an unpolarized BCS gas and a
polarized normal gas, as suggested in \cite{Bedaque2003a}. Such phase
diagrams have already been discussed in \cite{Koponen2007a} for a 3D
lattice, here we expand those results to a 1D system and discuss the
effects of Hartree and Gorkov corrections on the phase diagrams.

\begin{figure}
\centering
\includegraphics{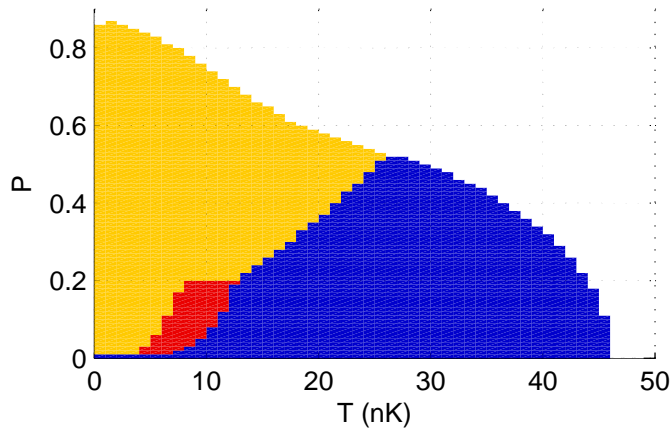}
\caption{The phase diagram of an imbalanced Fermi gas in an effectively 1D
  lattice. Colors: 
polarized superfluid $=$ blue (this is BCS at $P=0$), FFLO $=$ yellow,
  phase separation (PS) $=$ red, normal $=$ white. 
The average filling factor is $0.2$ atoms/lattice site in
each component, $J_x=0.07E_R$, and $U=-0.2 E_R$, where $E_R=\hbar^2k^2/2m$ is the
recoil energy. The FFLO area is remarkably large, with a high
critical polarization.}
\label{fig:1D_diagram_density}
\end{figure}

\begin{figure}
\centering
\includegraphics[width=0.495\textwidth]{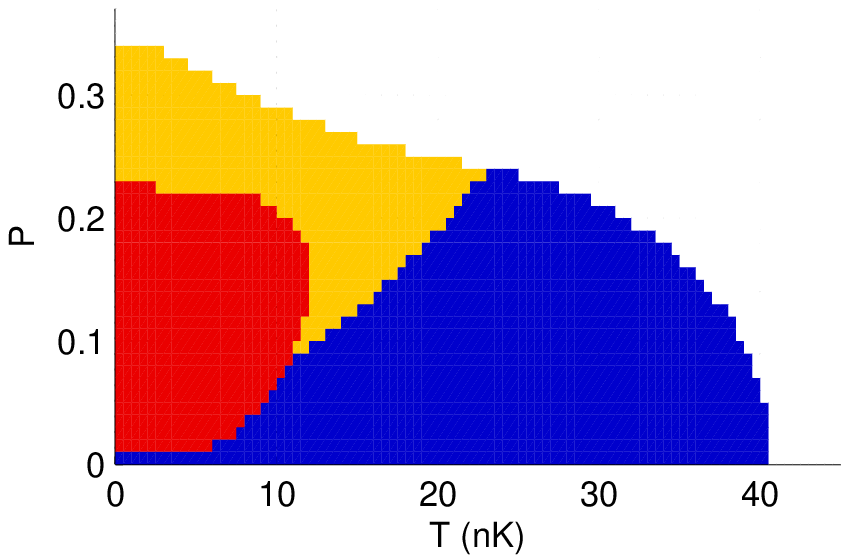}
\includegraphics[width=0.495\textwidth]{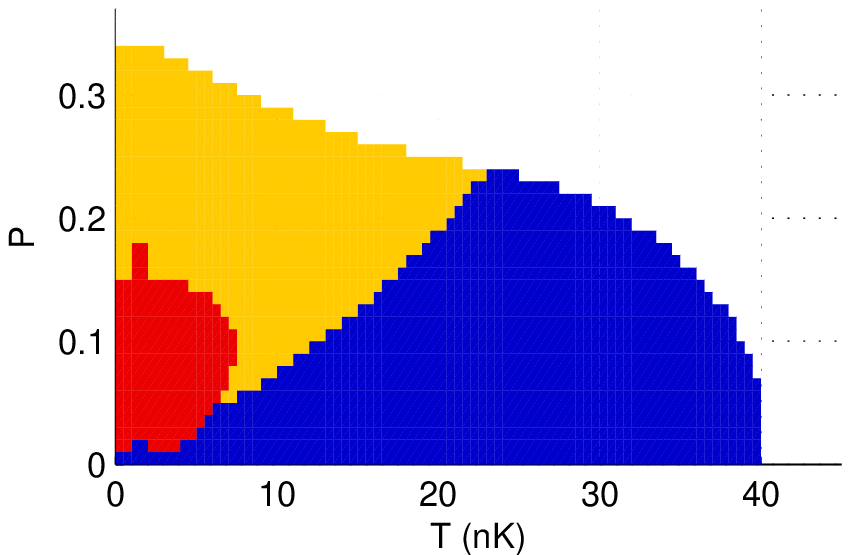}
\caption{Phase diagrams in a 3D lattice, without (left) and with
  (right) the Hartree corrections included. The colors are the same as
in figure \ref{fig:1D_diagram_density}: polarized superfluid $=$ blue,
FFLO $=$ yellow, phase separation (PS) $=$ red, normal $=$ white. The average filling 
is $0.2$ atoms/lattice site in each component, $J=0.07E_R$, and
$U=-0.26 E_R$. The phase diagram on the left is published in
\cite{Koponen2007a} and is shown here for comparison to the one on the
right.}
\label{fig:hartree_diagrams}
\end{figure}

The temperature-polarization phase diagram in a 1D lattice is shown in
figure \ref{fig:1D_diagram_density}. The interaction is chosen so that
the BCS critical temperature is close to that in the diagrams for the
3D lattice, shown in \ref{fig:hartree_diagrams}. These figures
demonstrate that with comparable conditions, a 1D lattice can support
much higher critical polarizations in FFLO and polarized superfluid
phases than the 3D system ($\sim 0.8$ vs. $\sim 0.3$ in zero
temperature with the chosen parameters).

The Hartree corrections rise from the Hartree terms, $U\langle
c\up[\vec{k}]^\dagger c\up[\vec{k}]\rangle c\down[\vec{k}]^\dagger
c\down[\vec{k}]$ and $U
c\up[\vec{k}]^\dagger c\up[\vec{k}]\langle c\down[\vec{k}]^\dagger
c\down[\vec{k}]\rangle$, which have been left out of the
Hamiltonian. When the density is constant in the system, the effect of
the Hartree terms is implicitly included in the number equations,
(\ref{eq:number_equations}). However, as the phase separated state
contains components with different densities \cite{Koponen2007a}, including the Hartree
terms explicitly may change the difference in free energies between
phase separation and e.g. FFLO. The free energy of system with
constant densities $n\up$ and $n\down$, with the Hartree terms included, is
\begin{eqnarray}
\nonumber F =& -\frac{\Delta^2}{U} + \mu\up n\up + \mu\down n\down + 2Un\up
n\down +\\ &\sum_{\vec{k}}\left(\xi\down[-\vec{k}+\vec{q}] +
  E_{-,\vec{k},\vec{q}} - \frac{1}{\beta }\ln \left(\left(1 + e^{-\beta
    E_{+,\vec{k},\vec{q}}}\right)\left(1 + e^{\beta
    E_{-,\vec{k},\vec{q}}} \right) \right)\right)
\end{eqnarray}

We have found that while including the
Hartree corrections brings the absolute difference, $F_{PS}-F_{FFLO}$
closer to zero, it enlarges the FFLO area in the phase diagram,
compared to phase separation. This is shown in figure
\ref{fig:hartree_diagrams}, where two phase diagrams with identical
conditions, except for the inclusion of the Hartree terms, for a 3D lattice,
are shown.

For a two-component Fermi gas with equal densities, the transition temperature
$T_{c\,0}$ to superfluid state has been calculated as
\begin{equation}
 T_{c\,0} \propto E_F \exp (1/N(0) U_0),
\end{equation}
where $N(0)$ is density of states at Fermi surface and $U_0$  is the two-body
interaction. This transition temperature is obtained by considering a purely two-body
interaction. However this picture can be modified due to the effect of
the medium.
This effect, which is also referred to as induced interaction \cite{Heiselberg2000a}, was
originally studied in 60s by Gorkov and Melik-Barkhudarov \cite{Gorkov1961a}. The main idea
goes back to the polarization in the medium by one fermion and its influence on a
another atom. 
The second atom is scattered by a modified total interaction,
including induced interaction, which can be written as
\begin{equation}
 U_{tot}=U_0+U_{ind}.
\end{equation}
Gorkov \textit{et al.} found that the
contribution of this effect reduces the transition temperature by a factor $\approx
2.2$. In our formalism, the interaction enters \textit{via} the first term in the grand
potential \ref{eq:Omega}. Therefore in principal one could take into account the induced
interaction correction by considering $U_{tot}$ instead of $U_0$. This requires to
calculate the induced interaction which is $\sim U_0^2$ and involves
the Lindhard function.

The induced interaction is \cite{Pethick2001a}
\begin{equation}
U_{ind} = U_0^2 L(\vec{k}) = U_0^2\int \frac{d\vec{p}}{(2\pi\hbar)^3} \frac{f_{\vec{p}}-f_{\vec{p}+\vec{k}}}{\epsilon_{\vec{p}+\vec{k}}-\epsilon_{\vec{p}}}
\end{equation}
Let us here qualitatively discuss the effect having
a lattice dispersion $\LD_{\vec{k}}$ instead of the usual quadratic dispersion
in free space. In the low density limit, the lattice dispersion becomes
effectively quadratic, and one can expect the same factor of $2.2$ reduction
of critical temperature as in the free space case \cite{Gorkov1961a}. The
term in the denominator of the
Lindhard function is, in case of the usual quadratic dispersion, of the
form $(\vec{k}+\vec{p})^2/2 - \vec{p}^2/2 = k^2/2 + kp \cos \theta$ (where $\theta$ is the angle
between $\vec{k}$ and $\vec{p}$). Around the point $k_i = \pi/2$ (for 1D lattice at half
filling), the lattice dispersions $1 - \cos (k_i)$ (here $i$ means $x$,$y$,$z$
depending on the dimensionality) become effectively linear. Linearizing
the lattice dispersion produces to the denominator of the Lindhard
function terms of the form $|k_i+p_i| - |p_i| = \sqrt{k_i^2 + p_i^2 +2k_i
p_i} - |p_i|$. Assuming that this can be approximated by a Taylor
expansion, one ends up with $k^2/2 + kp \cos \theta$ like in the quadratic
dispersion case. This indicates that the linear dispersion regimes might
not change the Gorkov correction considerably. However, only a numerical
evaluation of the integrals can give a definite answer. An interesting
issue is the behaviour of the $L(k)$ when the lattice filling is high enough
to include in $L(k)$ dispersions near the band edges, i.e. VanHove
singularities. Then, the dispersion $1 - \cos(k_i)$ around $k_i = \pi$ becomes
quadratic again, but with a negative effective mass. Naively, this means
the induced interactions would have opposite sign than usual, i.e. enhance
$T_C$ rather than suppress. However, the Lindhard function in this case, in
addition to quadratic dispersion with negative mass, contains also
additional terms due to expansion around $\pi$, and again a numerical
approach should be applied. Anyhow, one could still expect the Van Hove
singularity to reduce the effect of the Gorkov correction, as has been
shown in \cite{Orso2005a} for a 3D gas in a 1D lattice. Note that
this is not the same system as what we mean here by 1D
lattice, (which is a 1D gas in a 1D lattice) as explained in the
introduction. In our 1D lattice, the Van Hove singularity would not
influence the Gorkov correction since the singularity happens at full
filling, and that should be the same as low density limit due to symmetry
of pairing of particles and holes over half filling. However, in 2 and 3
dimensions the Van Hove singularities as well as regions of linear
dispersion affect $L(k)$ and it is an intriguing question how these effects
add up. Numerical investigation of these issues is a topic of our further
work.

\section{Conclusions}
We have considered phase diagrams of density imbalanced two-component
Fermi gases in optical lattices of dimension 1, 2 and 3. The phase
diagrams in the plane of the average chemical potential of the two
components, and the chemical potential difference, show striking effects
originating from the VanHove singularities of the lattice. These features
appear only for the FFLO state, not the BCS. Therefore they are unique
signatures of the FFLO state and reflect the fact that the nesting of the
Fermi surfaces in a lattice enhances FFLO pairing compared to case of the
homogeneous space. We show how these features preserve to finite
temperatures and finally disappear for very high temperatures.

Using LDA, we have demonstrated various shell structures that can appear
when the lattice is superimposed by a harmonic trapping potential. For the
studies of the FFLO state itself, it is useful that structures where only
FFLO and normal states appear can be found, but also more exotic shell
structures such as normal-FFLO-BCS-FFLO-normal are possible.

Density-density correlations are one possibility for observing the various
phases and states. We have shown here how the unpaired atoms in the FFLO
state leave a clear signature in the correlations. Especially in 1D the
signature is very prominent. This is true for the 1D gas in a 1D lattice
as considered here, as well as for the 1D continuum system (our system in
low density limit) which is also of high interest since the
one-dimensional confinement is known to enhance FFLO pairing
\cite{Hu2007a,Parish2007b} even without the lattice potential.

We also considered Hartree corrections which according to our calculations
tend to increase the FFLO area versus phase separation in the case of
fixed particle numbers. Outlook for further work includes more detailed
considerations of the Gorkov corrections, and studies of strongly
interacting gases instead of the weak and intermediate coupling regime
considered here.

\ack This work was supported by the National Graduate School in
Materials Physics, Academy of Finland (Projects
Nos. 115020, 213362, 121157, 207083) and conducted as a part of a
EURYI scheme award. See www.esf.org/euryi. 

\section*{References}

\bibliographystyle{h-physrev}
\bibliography{paperi}

\end{document}